\documentclass[manuscript]{aastex}

\shorttitle{Peculiar double minimum}
\shortauthors{Jain, Tripathy and Hill}

\begin{document}

\title{How peculiar was the recent extended minimum - A hint towards double minima}

\author{Kiran Jain, S. C. Tripathy and F. Hill}
\affil{National Solar Observatory, 950 N Cherry Ave., Tucson 85719, USA}

\begin{abstract}
In this paper, we address the controversy regarding the recent extended 
solar minimum as seen in helioseismic low- and intermediate-degree mode frequencies:
studies from 
different instruments identify different epochs of seismic minima. Here we use mode
frequencies from a network of six identical instruments,  Global Oscillation Network Group, 
continuously collecting data for more than 15 years, to investigate the epoch of minimum in 
solar oscillation frequencies  prior to the beginning of solar cycle 24. We include both
low- and intermediate-degree modes in the  $\ell$ range of 0 -- 120 and frequency
range of 2.0 -- 3.5 mHz. In this analysis, we demonstrate that there 
were indeed two minima in oscillation frequencies,
depending upon the degree of modes, or more precisely  the lower turning point
radius of the propagating wave. We also analyze frequencies as a function of
latitude to identify the beginning of solar cycle 24. We
observe two minima at high latitudes and a single minimum at mid/low latitudes. This scenario is 
in contrast to cycle 23 where the epoch 
of seismic minimum did not change with latitude or depth. Our results also
hint towards a possible role of the relic magnetic field  in
modifying the oscillation frequencies of modes sampling deeper layers.
\end{abstract}
\keywords{methods: data analysis --- Sun: helioseismology --- Sun: interior --- Sun: oscillations }

\section{INTRODUCTION}
The prolonged period of minimal solar activity and the delayed onset 
of solar cycle 24 have invoked a great deal of interest among solar
physicists to identify precisely when the activity minimum occurred and how different it
was from previously recorded minima. A variety of
data from the solar interior to the corona has been analyzed to 
seek the origin of such an unusually low solar activity for 
 a long duration (e.g., see articles in Cranmer, Hoeksema \& Kohl 2010). 
All of these analyses  indicate that the current minimum 
was much deeper than previous ones in modern era. Since the source of solar activity 
is believed to lie in a shear layer at the base of the convection zone, known as 
the tachocline, the analysis of helioseismic data is important in order to probe 
these regions.  Fortunately, the availability of continuous, consistent helioseismic 
data for two consecutive solar minima has provided a unique opportunity to study 
the changes in the solar interior that might have led to this unusual minimum. 

Using Birmingham Solar-Oscillation Network (BiSON) frequencies of 
low-angular degree ($\ell \le$ 3) modes 
for about three  solar cycles, \citet{bison09} found that the frequencies
were significantly lower in the extended minimum as compared to those 
during the previous ones. Similar results were obtained by \citet{david09} in 
an analysis based on the data collected by the space-based  Global Oscillations
at Low Frequencies (GOLF) on board the {\it Solar and Heliospheric Observatory (SOHO)}
spacecraft {\citep{gab95}}. They further pointed out that
cycle 24 started in late 2007, despite the absence of any visible activity 
on the solar surface. The analyses of modes in intermediate-degree 
 range  between 20 and 150, obtained from the Global Oscillation Network Group
  (GONG: \citet{harvey96}),  also showed lower oscillations frequencies during the
recent minimum \citep{jain10, sct10b}, however these studies did not identify the minimum
as occurring until the end of 2008, leading to an extended phase of low activity.
 Furthermore, the zonal and meridional flow patterns inferred from inverting frequencies also
suggested a delayed onset of the  new cycle. \citet{howe09} compared the evolution of 
the zonal flow pattern in the upper convection zone 
and  suggested that the mid-latitude flow
band corresponding to the recent cycle moved more slowly towards the equator 
than was observed in the previous cycle. As a result, there was a gradual
increase in the apparent length of the cycle. \citet{antia10} confirmed these
results and compared the behavior of the low-latitude prograde band
in past two minima. They pointed out that the flow band bifurcated in 2005 after merging in 2000, and
merged again in 2007 before disappearing around the beginning of the new activity cycle in 2009.
This is different from the previous minimum where split bands merged at the beginning of 
1996 and disappeared soon after the activity minimum was reached. Evidence that the
meridional circulation developed at medium-high latitudes before the beginning 
of a new cycle also supported the extended duration of low/no activity and delayed onset of
cycle 24 \citep{igh10}.

While most studies suggest that the beginning of cycle 24 occurred in late 
2008/early 2009, the results from the low-degree mode frequencies were puzzling.
In this context, we investigate the response of frequencies of low-degree as
well as intermediate-degree modes during the recent extended period of low activity. 
 The low-degree frequencies analyzed by \citet{david09}  were obtained by  
unresolved (Sun-as-a-star) Doppler velocity observations, 
while intermediate-degree frequencies, studied by \citet{sct10b}, 
were obtained with resolved  Doppler observations. 
To eliminate the influence, if any, of the techniques of observation and 
mode fitting, we use data from a network of six identical  instruments   
that treats both low- and intermediate-degree modes in a similar way. 
Finally, the depth and latitudinal dependence of the modes are studied during the last 
two minima. We present an analysis confined to modes that sense information from the 
tachocline region at different latitudes which provides useful information on the rise of a 
new cycle at mid latitudes.

\section{DATA}
 We use $p$-mode frequencies, obtained from the 
 GONG\footnote{ftp://gong.nso.edu/data/}, for the individual ($n, \ell, m$) 
multiplets, $\nu_{n \ell m}$, where $n$ is the radial order and $m$ is the
azimuthal order, running  from $-\ell$ to $+\ell$.  The mode frequencies for each 
multiplet were estimated from the $m-\nu$ power spectra constructed by the time 
series for individual GONG Month (GM = 36 days).  Here we used the standard GONG 
peak-fitting algorithm to compute the power spectra based on the multitaper spectral 
analysis coupled with a Fast Fourier transform \citep{rudi99}. Finally, Lorentzian 
profiles were used to fit the peaks in the $m-\nu$ spectra 
using a minimization scheme guided by an initial guess table \citep{ander90, hill98}. 
The guess table is based on fits to a special data set -- a 'grand average' 
of six 3-month periods at the beginning of GONG operation (R. Howe 2011, private 
communication) and is consistent for all data sets. 
 Frequencies for m $\ne$ 0 are calculated by applying a Legendre rotation 
expansion to the m $ = $ 0 frequencies.

The data analyzed here consist of 293 36-day overlapping data sets, with a spacing 
of 18 days between consecutive data sets,  covering the
period from 1995 May 7 to 2009 October 31 (i.e. GM = 1 -- 147)
 in 5-minute oscillation band;  2000 $\le \nu \le$ 3300 $\mu$Hz and  0 $\le \ell \le$ 120.
Thus, these data sets start about a year before the minimum of solar 
cycle 23 and end several months after the minimum
 of solar cycle 24. While most of earlier analysis are based
on the $m$-averaged frequencies, our analysis based on ($n, \ell, m$) multiplets 
has the advantage of studying latitudinal variations in frequency shifts that allows
us to investigate the changes around the activity belt where signatures of the rise of a 
new solar cycle first appear.

\section{RESULTS}
\subsection{Epochs of minima in acoustic oscillations and solar activity}
We calculate shifts in frequencies with respect to the guess
frequencies, %obtained from the model ``S" of \citet{smodel},
as used by the GONG pipeline in the fitting of a  particular multiplet 
($n, \ell, m$).  Since the frequency shift has a well-known dependency on  
frequency and mode inertia \citep{jain01}, we consider only those modes 
that are present in all data sets and the shifts are scaled by the mode inertia 
as defined by \citet{jcd91}.  The mean frequency shifts, $\delta\nu$, is 
calculated from the following relation,
\begin{equation}
\delta\nu(t) = \sum_{n \ell m}\frac{Q_{n \ell}}{\sigma_{n \ell m}^{2}}\delta\nu_{n \ell m}(t) / \sum_{n \ell m}\frac{Q_{n \ell}}{\sigma_{n \ell m}^{2}}
\end{equation}
where $Q_{n \ell}$ is the inertia ratio, $\sigma_{n \ell m}$ is the error
in frequency measurement, and $ \delta\nu_{n \ell m}(t)$ is the change in 
measured frequency for a given $n$, $\ell$  and $m$. We also calculate a 
11-point (i.e. 216 days) running mean of frequency shifts to smooth out short-term variations.
 This is shown by solid lines in all figures. 
%This conforms to the general practice of
%defining the epochs of solar minimum and maximum.
 The epoch of minimum in each case is defined by the lowest value in 
computed smoothed frequency shifts during the low-activity phase of
solar cycle. 

 The temporal variation of calculated $\delta\nu$ 
 is shown in Figure~1.  Dashed vertical lines correspond to the minima in
frequency shifts as determined from the smoothed values. It is evident that the 
minimum at the beginning of solar cycle 23 (henceforth {\it MIN23})
 occurred in early 1996 while the beginning of 2009 represented 
the minimum during the current extended phase of low activity. We also plot the 
variation of sunspot 
numbers\footnote{ftp://ftp.ngdc.noaa.gov/STP/SOLAR\_DATA/SUNSPOT\_NUMBERS/INTERNATIONAL} 
and the 10.7 cm radio 
flux\footnote{ftp://ftp.ngdc.noaa.gov/STP/SOLAR\_DATA/SOLAR\_RADIO/FLUX/Penticton\_Adjusted}, 
calculated over the same 
period as  the frequency shifts and smoothed with an 11-points running mean.
 The smoothed-sunspot numbers show  minima in mid 1996 and early 2009,
 respectively which are slightly displaced from the minima seen in radio 
flux.
 Furthermore, the frequencies during the minimum prior to solar cycle 24 
(henceforth {\it MIN24}) were lower than those during  {\it MIN23} 
in agreement with earlier results (\citet{sct10b}; and references therein). 
This difference is estimated to be around 0.015 $\mu$Hz.
The most significant discrepancy between  $\delta\nu$  and sunspot numbers 
is seen at the solar maximum where twin peaks show opposite behavior. 
The first peak in sunspot number (mid-2000) is higher than the secondary
pear (late-2001), while frequency shifts had a maximum around  secondary peak.
This has already been discussed, in both low-  and intermediate-degree modes  
\citep{wjc07,jain09}. The linear correlation coefficients calculated using 
unsmoothed and smoothed values of frequency shift and activity indices are given 
in Table~I. It is evident that there is a decrease in correlation during the 
minimum phase of the solar cycle as discussed in our earlier work with fewer 
data points \citep{jain09b}.  This decrease in correlation for smoothed values 
is not as significant as for the unsmoothed values of frequency shifts and 
activity indices. The correlation coefficients obtained for both indices using 
smoothed values are comparable, i.e. 0.89 and 0.91 for sunspot number and radio 
flux, respectively, at {\it MIN23}, and 0.96 and 0.97 at {\it MIN24}.
 We do not notice any significant change in correlation between
unsmoothed frequency shifts and activity indices at two minima, however smoothed
frequency shifts are better
correlated with solar activity than  at {\it MIN23}. Contrary to these results, 
by analyzing GOLF data, \citet{david09} found an anti-correlation between frequency 
shifts of $\ell$ = 0 and $\ell$ = 2,  and activity indices at {\it MIN24}.  In addition, 
\citet{bison09} also showed large difference between the activity proxy and the 
frequency shifts of low-$\ell$ modes, obtained from the BISON network, during the 
declining phase of the cycle 23.

\subsection{$\ell$ dependence of the epochs of minima}
To investigate the angular degree dependence on frequency shifts, we illustrate, 
in Figure~2, the epochs of minima in frequency shifts for different $\ell$ ranges, 
from 0 to 120 in steps of 10, at {\it MIN23} and {\it MIN24}. We define the epochs of
minimum as periods of lowest values of smoothed frequency shifts within 1$\sigma$ error. 
This criterion suggests a few epochs where the minimum is extended over more than one data point 
for a particular $\ell$ range.  It is seen that 
the frequency minimum at the {\it MIN23} for all $\ell$ ranges was in
early 1996 with the interval between the earliest and the latest minima is 
about 54 days. In contrast, there were two minima at {\it MIN24}
 depending on the $\ell$ range. For $\ell$ = 0~--~10, 10~--~20 and
20~--~30, the minimum was in late 2007 which coincides with the minimum seen
in GOLF low-$\ell$ frequencies \citep{david09}. On the other hand,
 the beginning of 2009 was the minimum for other $\ell$ ranges ($\ell \ge 30$)
 which supports the results of  intermediate-degree mode frequencies \citep{sct10b}. 
{Thus, for the first time,  the temporal variation of frequency shifts indicates an unusual 
scenario at {\it MIN24} where low- and intermediate-degree modes sense different
minima. }

To validate these findings, we divide the solar interior into four
distinct layers and  investigate the temporal variation with depth 
approximated by $\nu/\sqrt{l(l+1)}$, which is defined by the relation \citep{jcd91}
\begin{equation}
 r_t = \frac{c(r_t)}{2\pi}\frac{\sqrt{l(l+1)}}{\nu }  
\end{equation}
where $c$ is the sound speed  and  $r_t$  is the lower turning point radius. 
A higher value of  $\nu/\sqrt(l(l+1))$ denotes a smaller value of $r_t$ and 
hence a greater depth. In Figure~3, we plot temporal variation of frequency 
shifts for modes with their turning point radius in  the core ($r_t/R_0$ = 0.0 -- 0.3), 
the radiative zone ($r_t/R_0$ = 0.3 -- 0.7), near the tachocline ($r_t/R_0$ = 0.71 -- 0.73), 
and the convection zone ($r_t/R_0$ = 0.7 -- 1.0) where $R_0$ is the radius of the Sun. 
About 1\%, 12\% and 2\% of total modes are found to have
their turning points in the core, radiative zone and tachocline respectively, 
 while the rest of the modes were confined to the convection zone. 
Since the sound speed increases rapidly with depth, the 
acoustic modes spend less time in the deeper layers than the shallower layers, 
hence these modes are more influenced by changes in outer layers.
Figure~3 shows that the epoch of minimum at {\it MIN23} was at the beginning of 1996 in
all four cases while it varied with depth range at {\it MIN24}. Note that the modes
 with turning point radius in the core are mainly dominated by low-$\ell$ 
values and sense a minimum a year earlier than in other three layers. This is in 
agreement with Figure~2 of this paper and also with \citet{david09}. 
Further, we see two dips in frequency shifts in Panels (a) and (b) during the 
recent minimum but the lowest of these dips in Panel (a) is at the end of 2007 
while the minimum in Panel (b) is reached at two different times separated by about 
a year.  Two dips apart by 2 years during the minimum of cycle 24 have 
 been reported in low-$\ell$ BiSON and 
GOLF frequencies \citep{bison09,bison10}; these have  been 
interpreted as the signature of a second dynamo just below the solar 
surface.  We also notice the minimum in frequencies with $r_t$ around 
the tachocline (Panel c) and in the convection zone (Panel d)
 at the beginning of 2009  in agreement with the appearance of  activity
at the solar surface. These results clearly indicate the complex nature of the
relationship between oscillation frequencies and solar magnetic activity. 

\subsection{Latitudinal dependence of the epochs of minima}
Since the magnetic activity related to a new solar cycle emerges first at 
mid latitudes,  we follow the changes in mode frequencies as 
a function of the latitude using different values of $|m|/\ell$. For $|m|/\ell$ = 1, 
the modes are sensitive to the regions near the equator while $|m|/\ell$ = 0
 represents modes with sensitivity at higher latitudes. Figure~4 shows 
the mean variation in frequency shifts at selected values of $|m|/\ell$ during
{\it MIN23} and {\it MIN24}.  We again notice that the frequencies at all latitudes 
during {\it MIN24} are lower than those during {\it MIN23}. Further, in view of the above discussion, 
it is not surprising to obtain a single epoch of minimum at {\it MIN23}  
for all three values of $|m|/\ell$. On the  contrary, the frequency shifts 
hint towards a double minima at {\it MIN24} for  $|m|/\ell$ = 0.5, and a 
single minimum for $|m|/\ell$ = 0.7 and 0.9.  Although most of  well-known
surface activity indices do not indicate the minimum in late 2007, as the 
downward trend in their values continued until the end of 2008, there are 
observations where early signs of the beginning of the new cycle  were 
possibly seen. The appearance of a sunspot in Active Region 10981
 for three consecutive days during 2008 Jan 4 -- 6 at high latitude 
(30$^{\mathrm{o}}$ N) with new cycle  polarities hinted towards the rise 
of cycle 24  in early 2008 \citep{newcycle}. However, more sunspots fulfilling 
the criteria  of cycle 24 were not visible for several months after 2008 January. 
The emergence of three  big sunspots at low latitudes with previous cycle 
polarities a few months  later (magnetic polarity in accordance with cycle 23) 
suggested that the minimum after cycle 23 had  not yet been reached \citep{oldcycle}. 
These observations are not unusual as sunspots from both cycles are 
randomly seen during the minimum before approaching the bottom level of the 
activity.  The two strong dips seen in oscillation frequencies at certain 
latitudes can be understood in terms of localized changes
in the activity level. The mean variation in sunspot number
and radio flux during the two minima discussed here is shown in Figure~5. 
A close examination of the right panels of Figure~5 reveals that there was 
indeed a slight rise in activity at the beginning of 2008 but the trend did 
not continue. The two dips seen in oscillation frequencies may be interpreted 
as the manifestation of the competition between the magnetic fields from both 
the solar cycles. We also note periodic variations in $\delta\nu$ that are addressed 
in Section 4. 

\subsection{Minimum as seen in the tachocline region}
In order to explore the conditions near the tachocline at 
different latitudes during the extended minimum, we plot, 
in Figure~6, the mean variation in frequency shifts since 2004 for four
 $|m|/\ell$  values (0.5 $\le |m|/\ell \le $ 0.8) combined with 
lower turning point of modes  in the depth range of $r_t/R_0$ = 0.71 --0.73. 
Although there is a large scatter in $\delta\nu$, the 11-point 
running mean, in all cases, clearly indicates the strongest dips 
occurred around the end of 2008 except for  $|m|/\ell$ = 0.6 where 
another dip is seen in first half of 2008. Further, at $|m|/\ell$ = 0.5 and
0.8, we see a plateau from 2007 to mid 2008 and afterwards the frequency 
started to decrease again.  However, for $|m|/\ell$ = 0.7, which is closer to
the active-latitudinal belt, the frequencies increased at the beginning of 2007
 before reaching the minimum in last quarter of 2008.
This can be compared with the right two panels of Figure~5; although
there was a small increase  in solar activity around
the beginning of 2008, the activity level after
mid 2007 did not change much. This was the period of low
activity when a very few sunspots were visible for long periods,
and other solar activity indices, e.g. radio flux, followed a slow decrease
 before approaching the minimum. As the frequency variation for three 
values of $|m|/\ell$ values clearly shows a sharp dip near the minimum and an
increase in the frequencies after the minimum in late 2008, the variation at
$|m|/\ell$ = 0.6 is not conclusive where we see two minima; one in second quarter 
and other at the end of 2008.  The $|m|/\ell$ = 0.6 corresponds to the region near 
active-latitude belt where the signatures of a new cycle are expected to emerge first.
As mentioned earlier, the sunspots from both previous and new cycles  during the minimum phase 
appeared randomly on the solar disk and there were a large number of spotless days 
before the rise of cycle 24, we thus speculate that the different behaviour 
at $|m|/\ell$ = 0.6 might have resulted from these observed facts. 
The solar activity indicators, as shown 
in Figure~5, also support the rise of solar cycle 24 after 2008.

Figure~7 shows a similar variation in frequencies during the minimum of cycle 23. 
The top two panels, corresponding to $|m|/\ell$ = 0.5 and 0.6,
 show periodic variations while we find a  relatively smooth 
trend for the modes at $|m|/\ell$ = 0.7 and 0.8. As expected, for 
values of  $|m|/\ell$ lying within the active-latitudinal belt, the
variation is similar to that seen in activity indices (left panels of Figure~5).
Thus, the minimum reached in oscillation frequencies near the tachocline does not
change significantly with latitudes and coincides with the minimum
in visible surface activity.  

\section{DISCUSSION}
It is been argued for a long time that the perturbations of near-surface 
layers generated by the changes at the tachocline
 are mainly responsible for the changes in frequencies, 
however the observations during the extended minimum provides an indication 
that there might be some effects from layers as deep
as the core. The analysis of mode frequencies with lower turning points
in the inner 30\% of solar interior indicates a minimum earlier than
that obtained in outer layers or surface activity. Further, \citet{bison10} 
reported a quasi-biennial (2 year) signal in low-degree solar oscillation 
mode frequencies. Their study suggests that the 2 year signal, a 
predominantly additive contribution to the acoustic 11 year signal,
has its origin in significantly deeper layers than the 11 year signal and
is positioned below the upper turning point of the modes.  
Since the depth of a mode's upper turning point increases with decreasing 
frequency, one can expect low-$\ell$ modes to be more influenced by the shallower layers.
%As the depth of a mode's upper turning point increases with decreasing frequency
%and the frequency of the mode increases with depth, 
%one can expect low-$\ell$ modes to be more influenced by the shallower layers.}
Thus, they argued that a second dynamo, seated near the
bottom layer extending 5\% below the solar surface, may be responsible for
the 2 year signal.  Finally, the presence of two dynamos operating at
different depths might be responsible for the observed double minima during the
extended period of low activity.

To examine this quasi-biennial signal in the GONG frequencies, we follow the
procedure as adopted by \citet{bison10}; we subtract a smooth trend from mean 
shifts of independent time series by applying a boxcar filter of width 2 years. 
Figure~8 illustrates both the 11-year signal of the solar cycle and the frequency
residuals for four different layers discussed in Figure~3. Although we find a weak
2-year periodicity at high activity period which is similar to the periodicity
seen in proxies of solar activity \citep{beno98}, no significant trend during the 
periods of low activity is noticed. We further apply the fast Fourier 
transformation to identify any significant trend in the frequency residuals. We did not
find any prminent peak in this analysis which indicates that no other periodicity,
except the 11-year solar cycle signal, is present in the GONG data.
Thus, the absence of any significant trend of quasi-biennial signal (except during 
periods of high activity) suggests a different scenario that hints towards 
different physical processes which might be involved at different depths.

To investigate the solar origin of double minima seen in the oscillation frequencies, 
we consider two major fields inside the Sun: {\it (i)} a weak field at the core that 
plays a crucial role in the generation of 22-year magnetic polarity cycle \citep{mursula01} and is
believed to have been present in the Sun since its formation \citep{cowling45}, 
{\it (ii)} a megagauss field just beneath the convection zone that is responsible for the 
dynamo mechanism \citep{wad89} and the 11-year cyclic variation in solar activity. 
We speculate that a competition between these fields plays a significant role in explaining 
the variation in oscillation frequencies. 
 It is possible that the relic field senses the minimum eariler than
the field seated at the base of the convection zone. This is manifested in the form of early
minimum seen in low-degree modes whose lower tuning points lie in the core. Further,
the existence of two minima for the modes returning from radiative zone indicate an
inter-connection between weak and strong magnetic fields. We anticipate that these
fields affect the frequencies in all solar cycles but the influence due to relic field
is shielded by the higher field strength. However, during 
the extended minimum phase, the magnetic field generated by the solar dynamo was 
relatively weak and hence the effect of the relic field could be observed. 

The scenario presented here is speculative in nature since the information about the interior
can only be obtained through indirect measurements, e.g. frequencies of solar oscillations.
The continuing efforts to measure high-precision oscillation data for a complete 22-year magnetic cycle 
 may unveil the influence of a relic field on the variation of oscillation frequencies.

\section{SUMMARY}

Using uninterrupted and uniform acoustic mode oscillation frequencies 
from GONG, we investigated the variation of frequency shifts during the last 
two solar activity minima. Although the perturbations of near-surface 
layers generated by the changes at the tachocline are mainly responsible for 
the changes in frequencies, the observations during the extended minimum suggest 
that there might be some effect from the layers as deep as the core. Our analysis
provides evidence for a double minima in oscillation frequencies during the 
current prolonged low activity phase. It also supports previous results 
obtained with GOLF and GONG data for low- and intermediate degree modes respectively,
where  different epochs of minimum were reported on the basis of angular degree
\citep{david09,sct10b}.  In other words, the minima seen
in oscillation frequencies vary with the depth of turning point radius of the modes.
The waves reaching the inner 30\% of the interior exhibit a minimum one year earlier
than that from the outermost 30\% which is in agreement with the surface-activity minmum. 
Although there is considerable evidence for the variation of oscillation 
frequencies in phase with the surface activity, the analysis presented in this 
paper hints towards a possible role of relic magnetic fields in changing the 
oscillation frequencies which was addressed by \citet{mjt}. We also searched for a 
quasi-biennial signal in the GONG frequencies in order the explore the influence of 
a possible second dynamo in the shallower sub-surface layers. The unclear trend 
obtained in this analysis neither supports nor rules out the arguments made by 
\citet{bison10}. We emphasize that this analysis has been made possible with the 
access to continuous high-precision oscillation data for more than a solar cycle, 
in particular during the prolonged unusually low activity period. These results are 
important to understand the  origin of the modification of  oscillation frequencies 
which have been known to vary with the phase of activity cycle for last two decades, 
but whose detailed physical mechanism still remains
an open question.

\acknowledgments
We thank the anonymous referee for useful comments. We also thank John Leibacher for 
many discussions and critically reading the manuscript. This paper utilizes data obtained by 
the GONG program, managed by the National Solar Observatory, which is operated by AURA, Inc. 
under a cooperative agreement with the National Science Foundation. The data were acquired by 
instruments operated by the Big Bear Solar Observatory, High Altitude Observatory, Learmonth 
Solar Observatory, Udaipur Solar Observatory, Instituto de Astrof{\'i}sica de Canarias, and 
Cerro Tololo Interamerican Observatory. This work has been partially supported by NASA Grant 
NNG-08EI54I to National Solar Observatory.

\clearpage

\begin{deluxetable}
{lcccccccccccccccc}
\tabletypesize{\scriptsize} 
\tablecaption{Correlation coefficients between frequency shifts and activity indices
for three time samples. Shown here are the Pearson's linear coefficient ($r_P$), Spearman's rank
correlation ($r_S$) and the two-sided significance ($P_S$). }
\tablewidth{0pt}
\tablehead{
\colhead{Period}&\colhead{}&\multicolumn{7}{c}{Unsmoothed}&\colhead{}
&\multicolumn{7}{c}{Smoothed}\\
\cline{3-9} \cline{11-17}\\
 \colhead{}& \colhead{} & 
\multicolumn{3}{c}{Sunspot Number}& \colhead{}&\multicolumn{3}{c}{Radio Flux} &\colhead{}&
\multicolumn{3}{c}{Sunspot Number}& \colhead{}&\multicolumn{3}{c}{Radio Flux}\\
\cline{3-5} \cline{7-9} \cline{11-13} \cline{15-17}\\
\colhead{}& \colhead{} &\colhead{$r_P$}&\colhead{$r_S$}&\colhead{$P_S$}& \colhead{} &\colhead{$r_P$}&\colhead{$r_S$}&\colhead{$P_S$}
& \colhead{} &\colhead{$r_P$}&\colhead{$r_S$}&\colhead{$P_S$}& \colhead{} &\colhead{$r_P$}&\colhead{$r_S$}&\colhead{$P_S$}
}
\startdata
 All Data\tablenotemark{a} & &0.97 &0.97&0.0& & 0.98&0.98&0.0& &0.98&0.97&0.0& &0.99&0.99&0.0 \\
$MIN23$\tablenotemark{b} &  & 0.85 &0.82& 2$\times$10$^{-13}$& & 0.88 &0.86& 3$\times$10$^{-16}$& & 0.89 &0.87&1$\times$10$^{-15}$ & &0.91&0.90&3$\times$10$^{-21}$ \\
 $MIN24$\tablenotemark{c}  & &  0.86 &0.79& 1$\times$10$^{-17}$& & 0.89&0.84& 1$\times$10$^{-21}$& & 0.96  &0.92&7$\times$10$^{-31}$& &0.97&0.89&2$\times$10$^{-26}$ \\
\enddata 
\tablenotetext{a}{1995 May 7 -- 2009 October 31}
\tablenotetext{b}{1995 May 7 -- 1997 December 15}
\tablenotetext{c}{2005 December 28 -- 2009 October 31}
\end{deluxetable}

\clearpage
\begin{figure}
\epsscale{.80}
\plotone{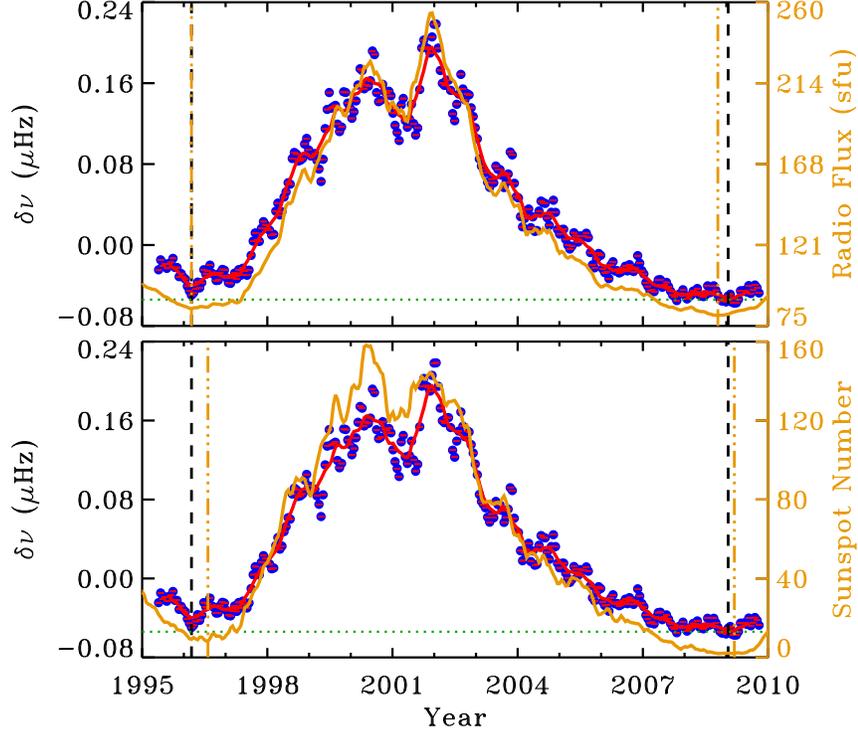}
\caption{Temporal evolution of frequency shifts (symbols) in the frequency range of 2000 
$\le \nu \le$ 3300 $\mu$Hz and  angular degree range of 0 $\le \ell \le$ 120. An 11-point
running mean of the frequency shifts is shown by the solid red line. The errors
in shifts are of the order of 10$^{-6}$ $\mu$Hz. The epochs of minima
in frequency shifts, based on the running mean,  are shown by 
vertical dashed black lines. The horizontal dotted green lines represent the lowest values 
of the running mean between cycles 23 and 24.  The variations in unscaled smoothed radio flux (top) and
sunspot numbers (bottom) are plotted with solid yellow lines, and the associated epochs of minima
in activity are shown by vertical dashed-dot-dot-dot yellow lines. 
\label{fig1}}
\end{figure}

\clearpage

\begin{figure}
\plotone{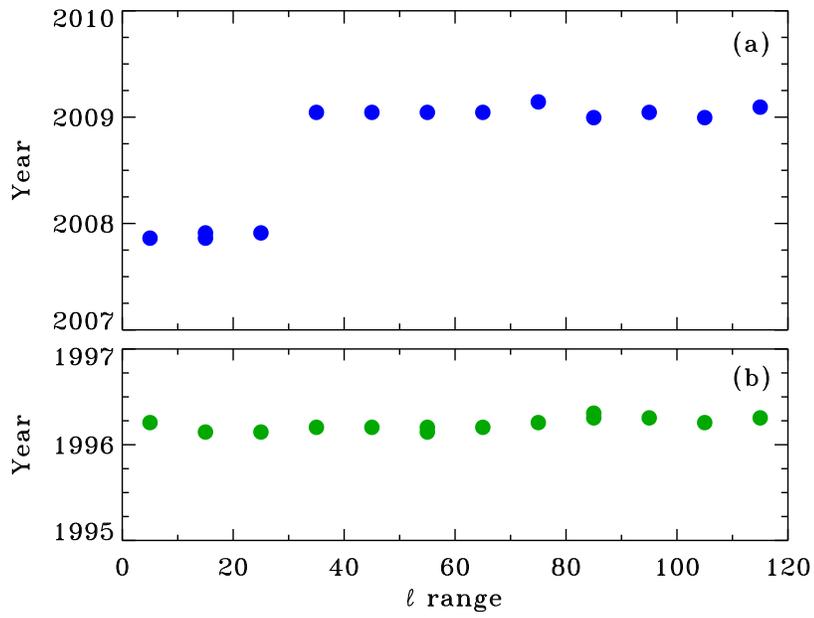}
\caption{Epoch of minimum for different $\ell$ ranges (a) before the beginning of
cycle 24, and (b) before the beginning of cycle 23 as determined by the 11-point
running mean of frequency shifts.
\label{fig2}}
\end{figure}

\clearpage

\begin{figure}
\plotone{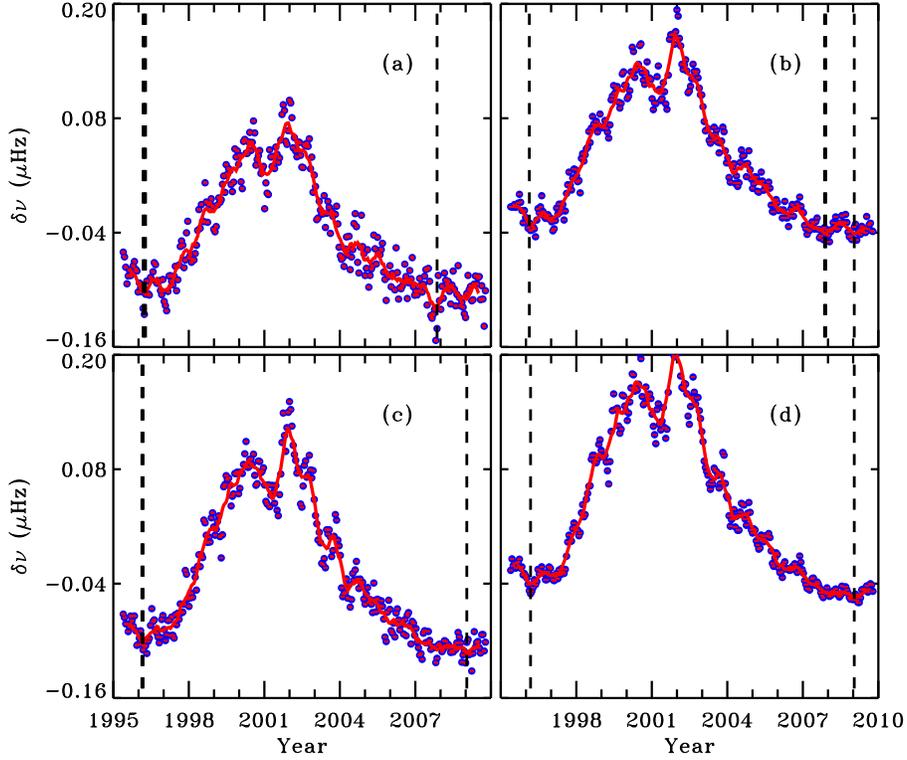}
\caption{Temporal evolution of the frequency shifts (symbols) for modes with turning
point radius in four different layers in the solar interior; (a) core ($r_t/R_0$ = 0.0 --0.3), 
(b) radiative zone ($r_t/R_0$ = 0.3 --0.7), (c) near tachocline ($r_t/R_0$ = 0.71 --0.73), and 
(d) convection zone ($r_t/R_0$ = 0.7 --1.0). The solid line represents an 11-point running mean 
of the  frequency shifts. Dashed vertical lines are drawn to identify the epochs of lowest values
 in frequency shifts.
\label{fig3}}
\end{figure}

\clearpage

\begin{figure}
\plotone{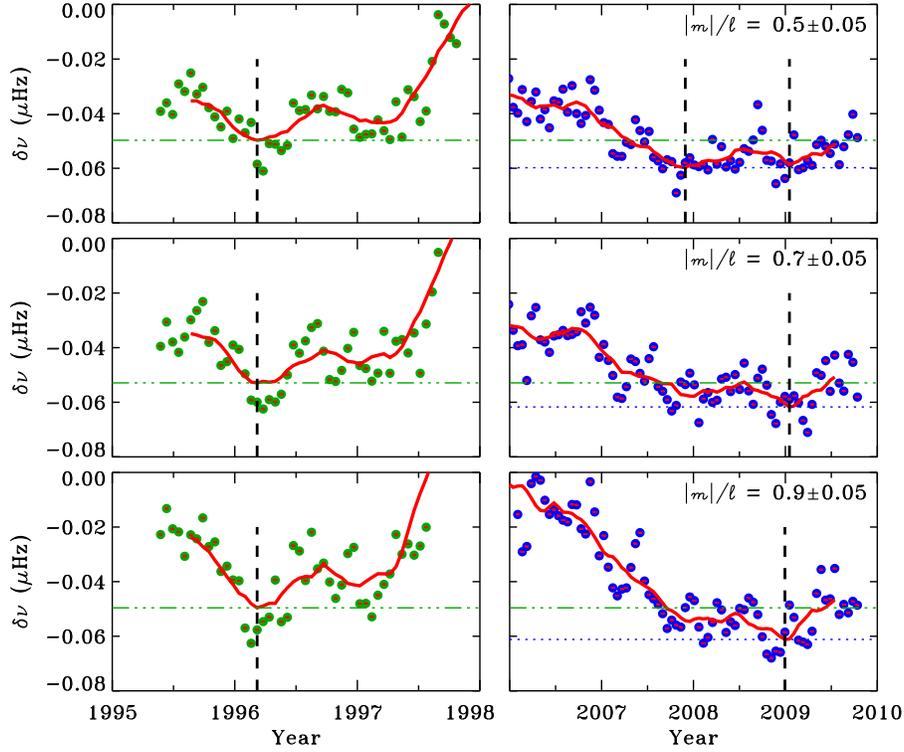}
\caption{Comparison of mean variation in frequency shifts between minima between
cycles 22 and 23 (left), and cycles 23 and 24 (right) at three different values of $|m|/\ell$. 
The solid line represents an 11-point running mean of the frequency shifts. The positions of 
minima are shown by the dashed vertical  lines. The dashed-dot-dot-dot and dotted  lines 
represent the lowest values of the running mean between cycles 22 and 23, and cycles 23 and 24, 
respectively. 
\label{fig4}}
\end{figure}

\clearpage

\begin{figure}
\plotone{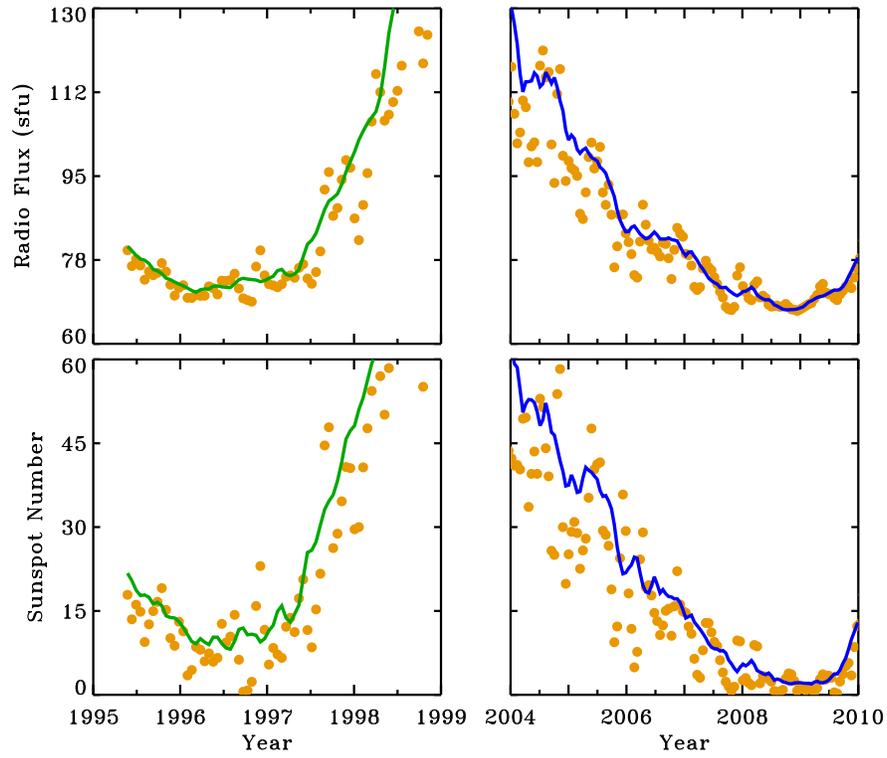}
\caption{ 
Symbols represent the mean variation in solar activity indices; (top) 10.7 cm radio flux, and
(bottom) international sunspot number during the periods of last two minima. An 11-point
running mean of  solar proxies is shown by the solid line.
\label{fig5}}
\end{figure}

\clearpage

\begin{figure}
\plotone{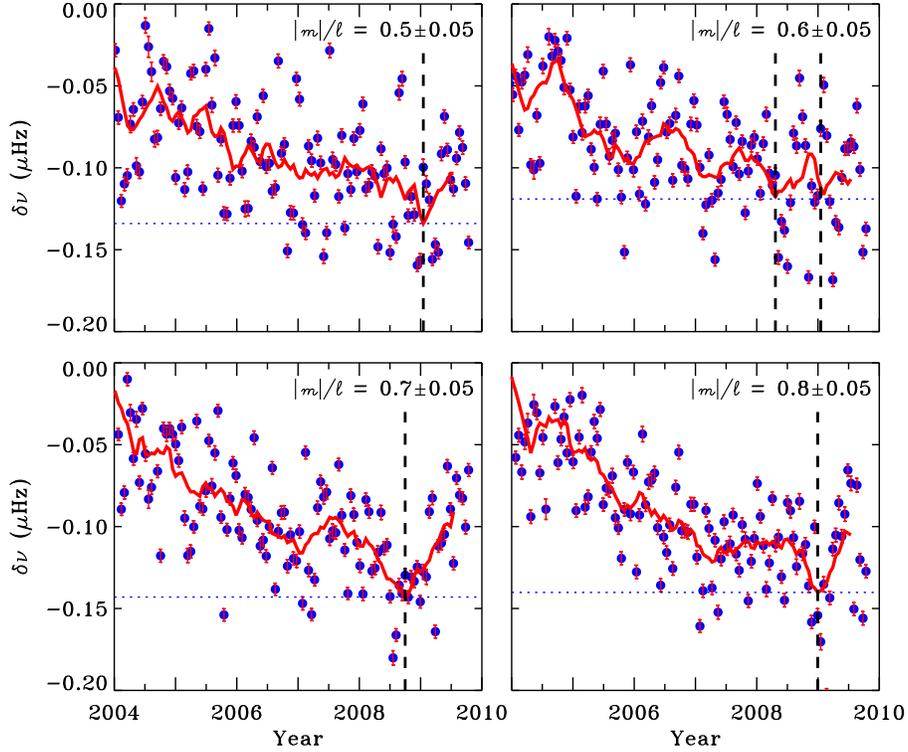}
\caption{Symbols represent the mean variation in frequency shifts for modes 
with turning points near the tachocline at four different values of $|m|/\ell$ 
during the minimum before solar cycle 24. The positions of minima are shown 
by the dashed vertical  lines. An 11-point running mean of the frequency 
shifts is shown by the solid line. The dotted  lines represent the lowest values 
of the running mean between cycles 23 and 24. 
\label{fig6}}
\end{figure}

\clearpage

\begin{figure}
\plotone{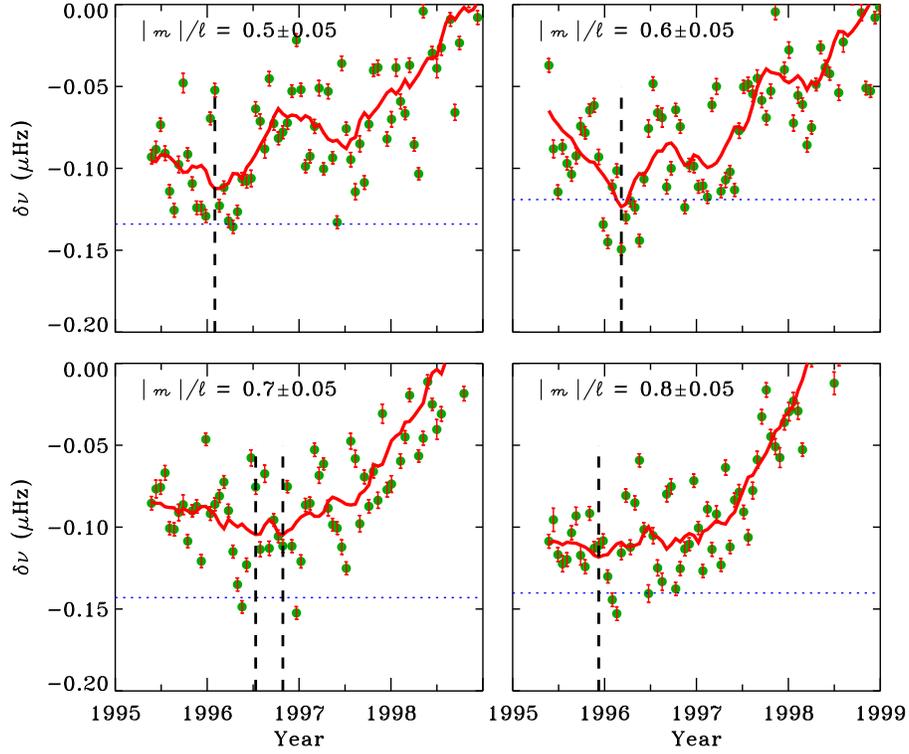}
\caption{The mean variation in frequency shifts (symbols) for modes with turning
points near the tachocline at four different 
values of $|m|/\ell$ during the minimum before solar cycle 23. 
The positions of minima are shown by the dashed
vertical  lines. The dotted  lines represent the lowest values 
of the running mean between cycles 23 and 24. 
\label{fig7}}
\end{figure}

\clearpage

\begin{figure}
\plotone{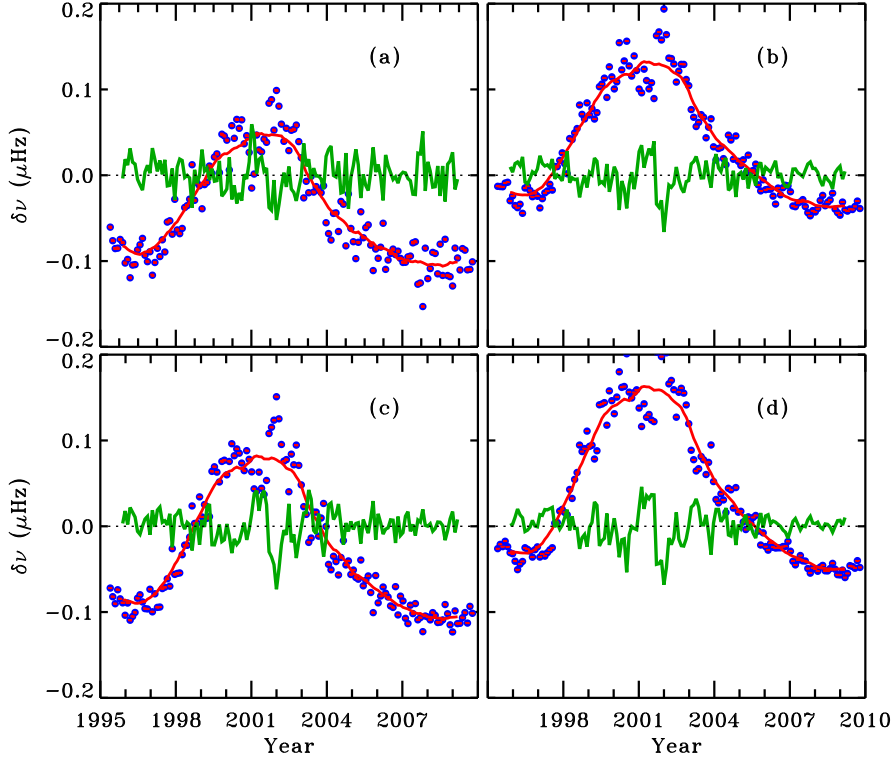}
\caption{
Temporal evolution of mean frequency shifts of independent time series 
(symbols) for modes with turning
point radius in four different layers in the solar interior; (a) core ($r_t/R_0$ = 0.0 --0.3), 
(b) radiative zone ($r_t/R_0$ = 0.3 --0.7), (c) near tachocline ($r_t/R_0$ = 0.71 --0.73), and 
(d) convection zone ($r_t/R_0$ = 0.7 --1.0). Solid red lines show the dominant 11 year signal of 
the solar cycle as calculated by applying a boxcar filter of the width of 2 year, 
and green lines are for the residual shifts after removing the 11 year signal.  
\label{fig8}}
\end{figure}

\end{document}